# High Bias Transport and Magnetometer Design in Open Quantum Dots

M. Switkes, A. G. Huibers, and C. M. Marcus

*Department of Physics, Stanford University, Stanford, California 94305*

K. Campman and A. C. Gossard

*Materials Department, University of California, Santa Barbara, California 93106*

We report transport measurements as a function of bias in open semiconductor quantum dots. These measurements are well described by an effective electron temperature derived from Joule heating at the point contacts and cooling by Wiedemann-Franz out-diffusion of thermal electrons. Using this model, we propose and analyze a quantum dot based sensor which measures absolute magnetic field at micron scales with a noise floor of $\sim 50 \; \mu\varphi_0/\sqrt{\text{Hz}}$ at 300 mK.

Quantum dots have been studied extensively as controllable systems in which quantum phenomena can be probed with transport measurements[1]. Most experiments are performed at low bias, typically less than 20 µV for dots with total conductance $g \geq e^2/h$ ("open" dots), however many proposed applications will require higher bias to produce easily measurable signals. Experiments in quantum dots in the tunneling regime ($g << e^2/h$) have exploited the nonlinear conduction of quantum dots at high bias for energy level spectroscopy[2]. In open dots, conduction remains essentially linear, and the primary effect of the increased bias is to heat the electrons in the dot. In a quantum dot sensor, the optimum bias balances the increased signal from high bias with the loss of quantum interference and increased noise from the elevated electron temperature. In this Letter, we present measurements of high bias transport in open quantum dots in thermal equilibrium, and show that the data are well described by an effective electron temperature. We then apply this effective temperature model to a quantum dot magnetometer, optimize its design parameters, and compare it to alternative technologies for micron scale magnetometry.

The measurements were made in a $^3$He cryostat over a temperature range 340 mK – 1.5 K with a quantum dot of area $A = 4.0 \; \mu\text{m}^2$ (Fig. 1 inset) defined by electron beam lithography on a GaAs/AlGaAs heterostructure using Cr/Au gates 160 nm above the two-dimensional electron gas (mobility $\mu = 1.2 \times 10^6$ cm$^2$/Vs and density $n_s = 1.8 \times 10^{11}$ cm$^{-2}$). We measured chordal conductance, $g = I_{bias}/V$, as a function of magnetic field $B$, cryostat temperature $T_0$, and RMS current bias $I_{bias}$, in two different configurations. The first configuration ("a" in Fig. 2 upper inset) used standard current-biased AC lock-in techniques at 200 Hz to measure $g$ at different cryostat temperatures with $I_{bias}$ fixed at 0.5 nA, small enough not to affect transport ($I_{bias} = 0.5$ and 1 nA give identical results at base temperature). The second configuration ("b" in Fig. 2 upper inset) measured chordal conductance for several values of $I_{bias}$ at $T_0 = 340$ and 580 mK using lock-in detection of a 43 Hz square wave excitation.

Figure 1 shows measurements of $g(B, T_0 = 340 \text{ mK}, I_{bias})$ along with the corresponding measurements of $g(B, T_0, I_{bias} = 0.5 \text{ nA})$. The close agreement of even the fine features of measurements made in these two configurations allows us to accurately assign an effective temperature, $T_{eff}$, to each bias current, shown in Fig. 2 for $T_0 = 340$ and 580 mK. This effective temperature can be interpreted as the electron temperature in the dot since, in the temperature and bias range of this experiment the primary thermalization mechanism, inelastic electron-electron collisions, occurs at a rate[3] $\tau_{ee}^{-1} = \left[\pi(k_B T)^2 / 4\hbar E_F\right] \ln(E_F / k_B T)$, which is comparable to or greater than the rate at which electrons escape from the dot, $\tau_D^{-1} = N\Delta/\pi\hbar$ where $N$ is the number of open channels in each point contact, $\Delta = 2\pi\hbar^2 / m^* A$ is the spin-degenerate mean single-particle level spacing, and $E_F$ is the Fermi energy. For our dot, $E_F = 6.3$ meV, $N = 1$, and $\Delta = 1.8$ $\mu$eV, giving $\tau_D = 1.2$ ns and $\tau_{ee} = 1.2$ ns at $T = 340$ mK and $\tau_{ee} = 0.17$ ns at $T = 1$ K. While $\tau_{ee} \sim \tau_D$ at base temperature, as the effective temperature increases, $\tau_{ee}$ decreases, allowing the electrons to thermalize in the dot over most of the experimental range. This is in contrast to recent experiments by Linke and coworkers[4] who found $T_{eff} = eV/k_B$ for electrons out of thermal equilibrium ($\tau_{ee} > \tau_D$), much higher than for thermalized electrons (Eq. 1)[5].

The measured effective temperature can be quantitatively modeled by balancing the energy added to the dot by high bias electrons with the energy lost to the environment. The power input is Joule heating from the point contacts, $P_{total} = I_{bias}^2 / g$. The heating occurs outside the point contact itself but not necessarily evenly split between the dot and the lead[6]; our data are best fit assuming ~10% asymmetry in heating. Cooling via the substrate is neglected since the electron-phonon scattering time[7] is much longer than $\tau_D$ for all temperatures studied here. The main cooling mechanism is Wiedemann-Franz out-diffusion of thermal electrons. Although originally developed for diffusive systems, the Wiedemann-Franz relation[8] has been predicted[9] and measured (to within a factor of two)[10] to apply to ballistic point contacts. For our geometry (electrical resistance in series, thermal resistance in parallel), the Wiedemann-Franz relation gives a thermal conductivity $\sigma_{th} = 4gL_0T$, with the Lorenz number $L_0 = (k_B/e)^2 \pi^2/3$. Assuming that the reservoirs are in good thermal contact with the cryostat, this yields

$$T_{eff}^2 = T_0^2 + \frac{P_{dot}}{2gL_0}. \qquad (1)$$

where $P_{dot}$ is the power heating the dot which, for a symmetric point contact, is half of the total Joule power. We find best agreement with our data (as shown in Fig. 2) for $P_{dot} = 0.4 P_{total}$. Presumably, the deviation of this coefficient from 1/2 is due to asymmetry in the point contacts[6].

The effective temperature model can be applied to predict the performance of a quantum dot magnetometer, which uses the large magnetoconductance of a quantum dot around $B = 0$ due to quantum-enhanced backscattering (weak localization) to measure absolute magnetic field with spatial resolution on the order of the dot size. The optimum operating bias balances the increased signal from higher bias with the loss of quantum mechanical coherence and increased noise due to higher $T_{eff}$, and is a function of other device parameters such as area and operating temperature. The voltage response of a quantum dot to magnetic field, $\chi \equiv dV/dB = -(I_{bias}/g^2)dg/dB$, can be calculated for a chaotically shaped ballistic dot from the Lorentzian lineshape of the weak localization[11] $\langle g(B) \rangle = \langle g \rangle_{B \neq 0} - \langle \delta g \rangle / (1 + 4B^2/B_c^2)$, whose height[12] and width[13] are respectively $\langle \delta g \rangle \equiv \langle g \rangle_{B \neq 0} - \langle g \rangle_{B=0} \sim (e^2/h) N/(2N + \gamma_\phi)$, and $B_c = \varphi_0^{(n)} \kappa^{1/2} (2N + \gamma_\phi)^{1/2}$, where $\langle g \rangle_{B \neq 0} = (e^2/h)N$ is the conductance at large magnetic field, $\varphi_0^{(n)} = 2\varphi_0 = h/e$ is the normal state magnetic flux quantum, $\kappa \propto A^{-5/2}$ is a constant of order 1 dependent on dot geometry, and $\langle \cdot \rangle$ indicates an ensemble-averaged quantity[14]. Both $\langle \delta g \rangle$ and $B_c$ depend on temperature through dephasing[15], which we have expressed in terms of the dimensionless rate $\gamma_\phi = 2\pi\hbar/(\Delta \tau_\phi)$. The low temperature (340 mK $\leq T \leq$ 4 K) dephasing time, $\tau_\phi$, has recently been measured in similar devices for a range of dot areas from 0.4 to 4 $\mu m^2$ with[16] $\tau_\phi \approx [1.4T^2 \ln(73/T) + 13T]^{-1}$ for $\tau_\phi$ in ns and $T$ in K. Combining these equations and replacing $T$ with $T_{eff}$ yields the theoretical average voltage response $\langle \chi(B, T_0, I_{bias}) \rangle$ which has a maximum in magnetic field $\chi_{max} \equiv \chi(B_{max})$ at $B_{max} = B_c(1 - \langle \delta g \rangle / \langle g \rangle)^{1/2}/2\sqrt{3}$. Two measured values of $\chi_{max}(T_0 = 340 \text{ mK}, I_{bias})$ are plotted in the lower inset of Fig. 2 along with the theoretical ensemble averaged value derived above, $\langle \chi_{max}(T_0 = 340 \text{ mK}, I_{bias}) \rangle$. The measured values differ from the ensemble averaged value by as much as a factor of two due to universal conductance fluctuations (UCF) which remain in the *unaveraged* measurements[16,17]. In principle, UCF can be used to improve the response of the magnetometer by "tuning" the dot with shape distortion gates[14] to a configuration with a high $\chi_{max}$. Figure 3a shows the theoretical $\langle \chi_{max} \rangle$ as a function of dot area and temperature for the bias (shown in Fig. 3b for $T_0 = 300$ mK) which yields the highest $\langle \chi_{max} \rangle$. Note that this is the average voltage response which can be improved by tuning the dot to take advantage of the UCF.

The fundamental limit on the magnetic field resolution of the quantum dot is set by intrinsic RMS voltage noise, $S_V^{1/2}$, which gives rise to a magnetic field noise $S_B^{1/2} = S_V^{1/2}/\langle \chi \rangle$. Shot noise and Johnson-Nyquist noise combine to give $S_V^{1/2} = \sqrt{4k_B T^*/g}$ with[18] $T^* = T[1 + \alpha(eI_{bias}/2gk_B T)\coth(eI_{bias}/2gk_B T) - \alpha]$ where $\alpha = 1/4$ is the shot noise suppression factor in quantum dots[19]. Figure 3c shows this noise as a function of dot area and cryostat temperature for the bias (shown in Fig. 3b for $T_0 = 300$ mK) which minimizes the noise. Again this is the average noise which can be decreased by tuning the dot.

The noise in quantum dots compares favorably to other technologies, such as sub-micron ballistic Hall probes[20], which also measure absolute magnetic field. For a 1 $\mu m^2$ device at 300 mK, a quantum dot has a flux noise $S_\varphi^{1/2} = A S_B^{1/2} \sim 50$ $\mu\varphi_0/\sqrt{\text{Hz}}$ (Fig. 3d), compared to $\sim 80$ $\mu\varphi_0/\sqrt{Hz}$ for the submicron Hall probe[21]. Both of these devices have an order of magnitude higher flux noise than a state of the art commercial SQUID[22], $\sim 3$ $\mu\varphi_0/\sqrt{\text{Hz}}$. Note however that the ultimate size limit for SQUIDs is the magnetic penetration length (~85 nm in Nb), whereas for devices based on chaotic quantum interference, the ultimate size limit is the Fermi wavelength which can be orders of magnitude smaller.

In conclusion, we have measured the effects of high bias on transport in open quantum dots and find that an effective temperature model, balancing Joule heating and Wiedemann-Franz heat conduction, fits the data well. We use this model to analyze a magnetometer based on quantum dot technology and compare it to alternative technologies.

We thank Dan Prober, Laurens Molenkamp, and Andrei Geim for very useful discussions. We gratefully acknowledge support at Stanford from the Office of Naval Research YIP program under Grant N00014-94-1-0622, the Army Research Office under Grant DAAH04-95-1-0331, the NSF-NYI program, the A. P. Sloan Foundation, and support for AGH from the John and Fannie Hertz Foundation. We also acknowledge support at UCSB by the AFOSR under Grant 49620-94-1-0158.

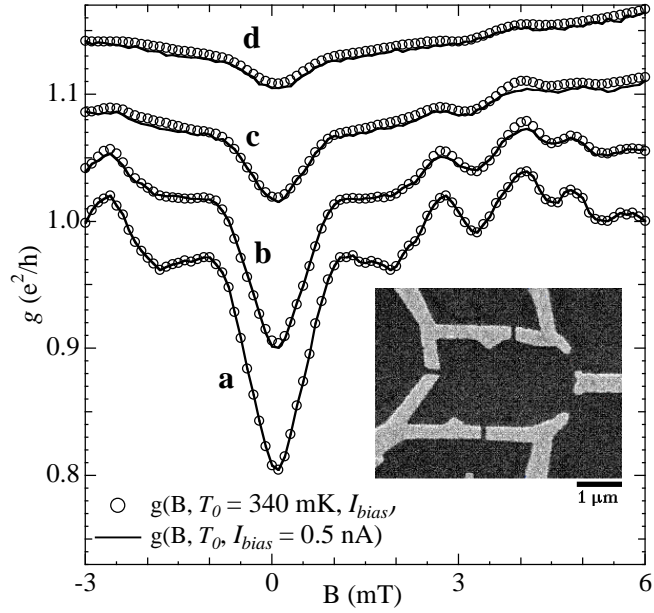

**Fig 1.** Magnetoconductance, $g(B, T_0 = 340 \text{ mK}, I_{bias})$, for $I_{bias}$ = 0.5 (a), 3 (b), 7 (c), and 12 nA (d) (open circles), with corresponding $g(B, T_0, I_{bias} = 0.5 \text{ nA})$ for $T_0$ = 340 (a), 410 (b), 680 (c), and 990 mK (d) (solid lines) showing the equivalence between bias and temperature. Each pair of curves has been offset for clarity. Inset: Micrograph of measured device.

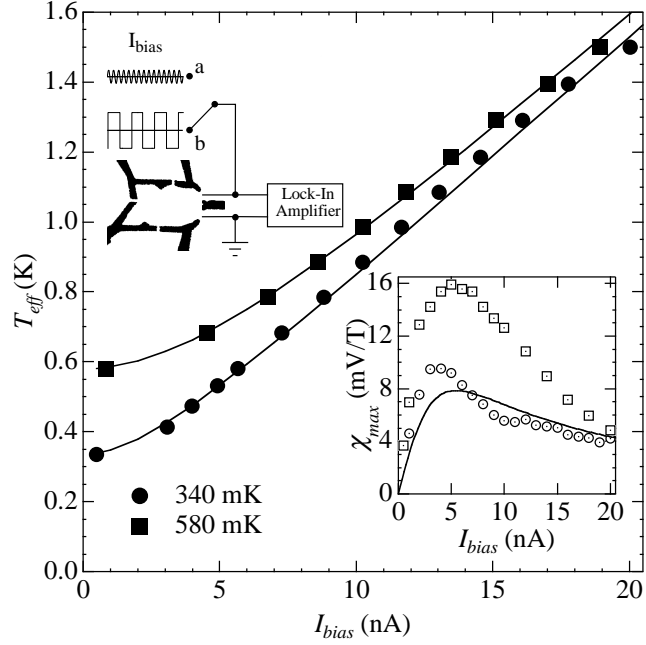

**Fig 2.** Effective temperature, $T_{eff}$ as a function of $I_{bias}$ for $T_0 = 340$ and 580 mK (filled symbols) along with the theoretical value from Eq. (1) obtained by balancing Joule heating and Wiedemann-Franz cooling (solid lines). Upper Inset: Schematic of the measurement circuit for (a) low $I_{bias}$ and (b) variable $I_{bias}$ measurements. Lower Inset: Two measurements of the voltage response $\chi_{max}$ at $T_0 = 340$ mK (open symbols) along with the theoretical value for the *ensemble-averaged* voltage response $\langle \chi_{max} \rangle$ (solid line) which differs from measured values due to UCF.

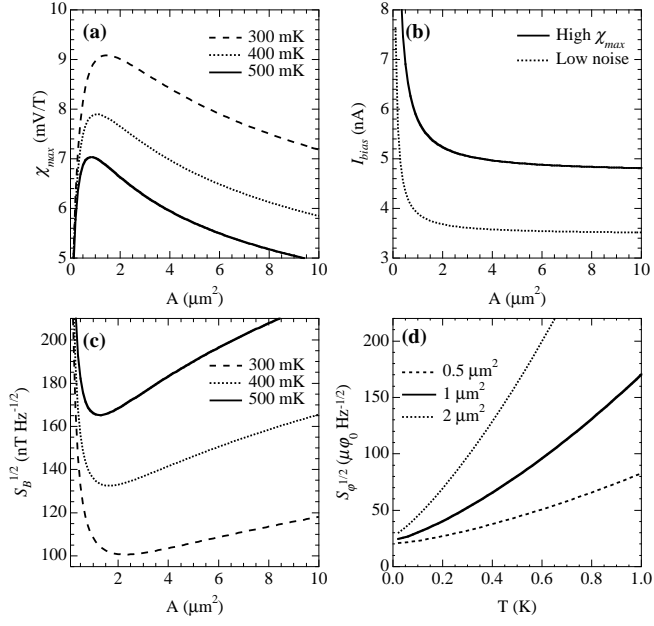

**Fig. 3** Theoretical performance characteristics of the quantum dot magnetometer based on $T_{eff}$ model: (a) Maximum voltage response $\langle \chi_{max} \rangle$ as a function of dot area and temperature with $I_{bias}$ chosen for greatest voltage response. (b) Bias current necessary for best voltage response (solid line) and lowest noise (broken line) at 300 mK. (c) Magnetic field noise $S_B^{1/2}$ with $I_{bias}$ as shown in Fig. 3(b). (d) Magnetic flux noise $S_\varphi^{1/2} = A S_B^{1/2}$.